\documentclass[aps,prb,onecolumn,superscriptaddress,floatfix]{revtex4}

\usepackage{graphicx}
\usepackage{dcolumn}
\usepackage{bm}
\usepackage{float}
\usepackage{amsmath}
\usepackage[latin1]{inputenc}
\begin{document}

\title {Spatially modulated magnetic structure of EuS due to the tetragonal domain structure of SrTiO$_3$}

\author{Aaron J. Rosenberg}
\affiliation{Department of Applied Physics, Stanford University, Stanford, CA 94305, USA}

\author{Ferhat Katmis}
\affiliation{Fracsis Bitter Magnetic Lab, Massachusetts Institute of Technology, 77 Massachusetts Avenue, Cambridge, Massachusetts 02139, USA}
\affiliation{Department of Physics, Massachusetts Institute of Technology, 77 Massachusetts Avenue, Cambridge, Massachusetts 02139, USA}

\author{John R. Kirtley}
\affiliation{Department of Applied Physics, Stanford University, Stanford, CA 94305, USA}

\author{Nuh Gedik}
\affiliation{Department of Physics, Massachusetts Institute of Technology, 77 Massachusetts Avenue, Cambridge, Massachusetts 02139, USA}

\author{Jagadeesh S. Moodera}
\affiliation{Fracsis Bitter Magnetic Lab, Massachusetts Institute of Technology, 77 Massachusetts Avenue, Cambridge, Massachusetts 02139, USA}
\affiliation{Department of Physics, Massachusetts Institute of Technology, 77 Massachusetts Avenue, Cambridge, Massachusetts 02139, USA}

\author{Kathryn A. Moler}
\affiliation{Department of Applied Physics, Stanford University, Stanford, CA 94305, USA}
\affiliation{Department of Physics, Stanford University, Stanford, California 94305, USA}
\affiliation{Stanford Institute for Materials and Energy Sciences, SLAC National Accelerator Laboratory, 2575 Sand Hill Road, Menlo Park, California 94025, USA}

\begin{abstract}
The combination of ferromagnets with topological superconductors or insulators allows for new phases of matter that support excitations such as chiral edge modes and Majorana fermions. EuS, a wide-band-gap ferromagnetic insulator with a Curie temperature around 16 K, and  SrTiO$_3$ (STO), an important substrate for engineering heterostructures, may support these phases. We present scanning superconducting quantum interference device (SQUID) measurements of EuS grown epitaxially on STO that reveal micron-scale variations in ferromagnetism and paramagnetism. These variations are oriented along the STO crystal axes and only change their configuration upon thermal cycling above the STO cubic-to-tetragonal structural transition temperature at 105 K, indicating that the observed magnetic features are due to coupling between EuS and the STO tetragonal structure. We speculate that the STO tetragonal distortions may strain the EuS, altering the magnetic anisotropy on a micron-scale. This result demonstrates that local variation in the induced magnetic order from EuS grown on STO needs to be considered when engineering new phases of matter that require spatially homogeneous exchange. 
\end{abstract}

\maketitle

EuS is a well-studied wide-band-gap ferromagnetic insulator with a NaCl-type structure (lattice constant of 5.94 {\AA}) and a bulk Curie temperature of 16.8 K \cite{mauger1986magnetic,idzuchi2014critical,kotzler1986change}. It has historically been used as an efficient spin filter to spin-polarize charge currents. \cite{nagahama2007enhanced,leclair2002large,miao2009magnetoresistance,moodera2007phenomena,hao1990spin,moodera1988electron}. Because EuS is considered a simple Heisenberg ferromagnet (a ferromagnet that can orient an any 3D direction \cite{bohn1984spin}), it has long been considered a model system to test theories of magnetism \cite{passell1976neutron,als1971critical,als1976neutron,dietrich1976neutron}. EuS has been of recent interest because it may induce magnetic order in topologically non-trivial systems \cite{wei2013exchange, lee2016optic}. For example, 3D topologically insulating Bi$_2$Se$_3$ \cite{zhang2009topological,zhang2010crossover,bianchi2010coexistence,kim2012surface,zhang2010first} has been combined with EuS to induce high-temperature ferromagnetism in the Bi$_2$Se$_3$ \cite{katmis2016high}. Additional potential applications include the creation of topological superconductivity to produce zero-energy Majorana fermion modes \cite{hasan2010colloquium, sau2010generic}, the topological magneto-electric effect \cite{qi2008topological, essin2009magnetoelectric}, a magnetic monopole \cite{qi2009inducing}, and the quantum anomalous Hall effect \cite{laughlin1983anomalous,chang2013experimental}. In the latter example, one could grow a heterostructure of EuS/topological insulator/EuS with the intention of breaking time reversal symmetry on the top and bottom surface states in order to observe chiral edge modes \cite{qi2008topological}.

SrTiO$_3$ (STO) is a common substrate for growing new heterostructures such as high-temperature superconductors \cite{qing2012interface, wu1987epitaxial, ge2015superconductivity}, ferroelectrics \cite{sun2004evolution,jiang1999abrupt}, and electronic systems with high spin-orbit coupling \cite{caviglia2010tunable}. STO is a perovskite band insulator with a cubic unit cell. Excitingly, it becomes an unconventional superconductor when doped \cite{klimin2012microscopic, koonce1967superconducting}, and the interface between STO and another perovskite band insulator, LaAlO$_3$, is both conducting \cite{ohtomo2004high,thiel2006tunable} and superconducting \cite{reyren2007superconducting}. At 105 K, STO undergoes a cubic-to-tetragonal structural phase transition because of small rotations of the Ti-O octahedra that causes the unit cell to elongate along one of the crystallographic axes \cite{cowley1964lattice}. Without external strain, the STO unit cell can elongate along any of the original cubic axes forming structural domains separated by twin planes. In terms of the original cubic directions, the twin planes are along (110)$_p$, (101)$_p$, and (011)$_p$. Recent studies have shown that the low-temperature twin structure affects both the interfacial conductivity in LaAlO$_3$/SrTiO$_3$ heterostructures \cite{kalisky2013locally,honig2013local} and the superconducting transition temperature in STO \cite{noad2016variation}.

We measured the magnetic spatial landscape in four EuS/STO-based heterostructures using a scanning superconducting quantum interference device (SQUID) susceptometer in a $^4$He cryostat. Unless otherwise indicated, the data in this manuscript were taken on a thin film (5 nm) of (001)$_p$-oriented EuS grown on a (001)$_p$-oriented STO substrate, but we observed similar effects in samples with (110)$_p$- and (111)$_p$-oriented STO substrates, and samples with 3 nm thick EuS (see appendix A).

The SQUID sensor measures the total flux through the pickup loop, which is integrated with the body of the SQUID through well-shielded superconducting coaxial leads. The pickup loop size and height above the surface determine the spatial resolution. The pickup loop has an inner radius of 1 $\mu$m and an outer radius of 1.5 $\mu$m, resulting in an effective radius of $1.24  \mu$m \cite{brandt2005thin}. The pickup loop is centered in a single-turn field coil with a 2.5 $\mu$m inner radius that can be used to apply a local magnetic field to the sample. Using this sensor, we  simultaneously probed the static ferromagnetism (magnetometry) as well as the susceptibility (susceptometry) of the sample \cite{kirtley2016scanning}. Magnetometry imaging was carried out by measuring the magnetic flux through the SQUID pickup loop, which is the z-component of the magnetic field produced by the sample convolved with the pickup loop's point-spread function. Susceptometry involved applying a small alternating current (ac) current to the field coil and recording the flux through the pickup loop using standard lock-in techniques \cite{huber2008gradiometric}. To image, we fixed the SQUID sensor above the sample and rastered the sample in the x-y plane using an attocube piezoelectric stack. We thermally coupled the SQUID sensor directly to the liquid helium bath, and we thermally isolated the sample and heater, allowing us to study the magnetic behavior of the sample even at temperatures higher than the superconducting transition temperature of the SQUID (T$_C$ = 9 K).

\begin{figure}
\centering
\includegraphics[scale=.36]{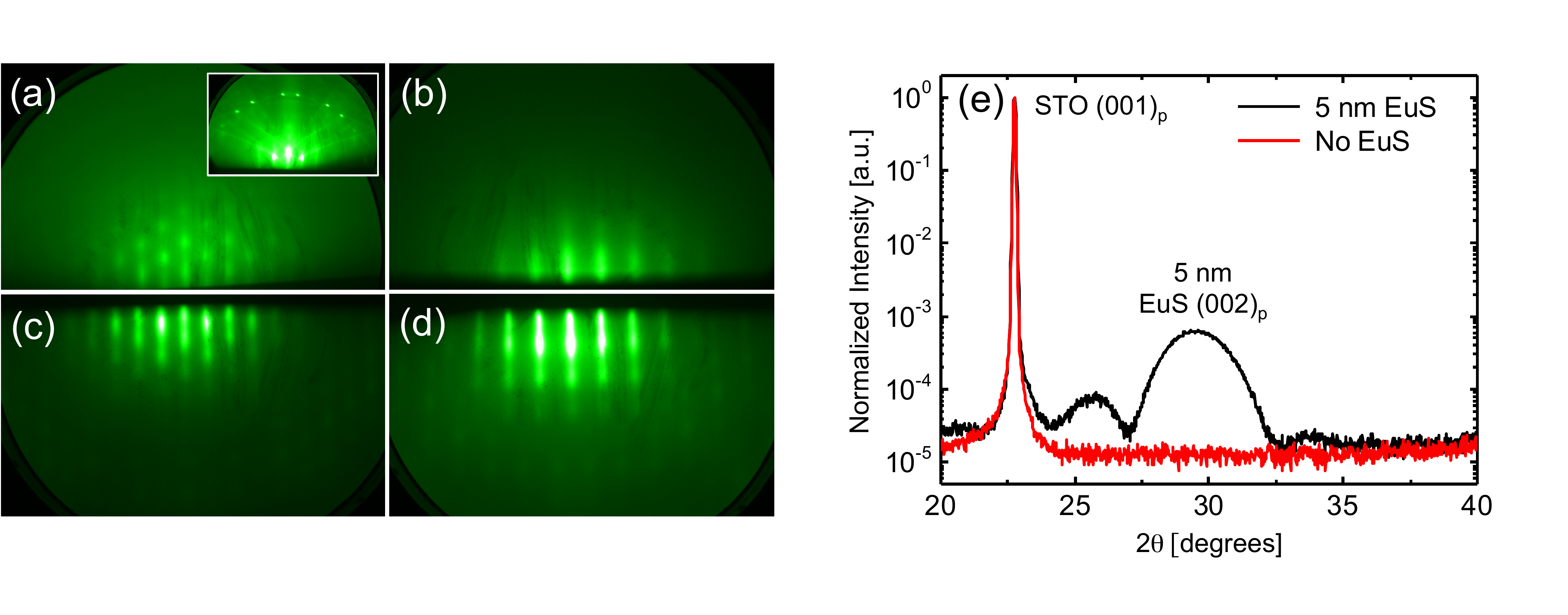}
\caption{Captured RHEED patterns of EuS after growth and annealing are shown in (a)-(d). RHEED snap-shot taken with 15 keV electron beam along the [110]$_p$-azimuth (a) and [001]$_p$-azimuth (b) after growth and annealing, respectively, (c) and (d). (e) X-ray diffraction of the EuS/STO heterostructure (black line) and STO substrate without EuS layer (red line). Besides the characteristic intense substrate reflection at the low angle side, another reflection is visible around $\simeq$ 29.5$^\circ$, which belongs to the (002)$_p$ reflection of EuS. The representative data also shows pronounced Laue oscillations (oscillations in the vicinity of the EuS Bragg peak), which indicate the coherency of the top and the bottom surface parallelism of the epitaxial EuS layer.}
\label{fig:EuS_rheed_xray}
\end{figure}

Epitaxial EuS growth and STO substrate preparation were performed in a custom-built molecular beam epitaxy system under base pressure of $2 \times 10^{-10}$ Torr. The system is equipped with ultra high-purity source materials for \textit{in situ} growth and protection of the films, as well as units to monitor the thickness of the layer during growth. The interface formation and structural evolution of the grown layer were displayed via an \textit{in situ} reflection high-energy electron diffraction (RHEED) apparatus. The STO epi-ready substrate was prepared in situ after several heat treatments to form an atomically flat surface, which was ensured via RHEED (Fig. \ref{fig:EuS_rheed_xray}(a) inset).

Due to the high reactivity of europium atoms and dissociation problems with sulfur, the EuS was evaporated congruently from a single electron-beam source. To avoid kinetic surface roughening, a $\simeq$ 1 - 3 {\AA} min$^{-1}$ growth rate was used to produce a quasi-smooth surface at 523 - 563 K. Layers were grown at 523 K and annealed after growth at 563 K until the layer quality was optimized, as determined via analysis of the RHEED pattern (Fig. \ref{fig:EuS_rheed_xray}(a), (b)). Even high-temperature growth does not provide sufficient surface mobility to the EuS molecules; therefore, surface roughening occurs above a critical thickness of 3-4 nm \cite{wei2013exchange,katmis2016high}. While annealing the film after the growth (Fig. \ref{fig:EuS_rheed_xray}(c), (d)), quasi-2D layer streaks transformed into mostly 2D modes, indicating surface smoothing after annealing, which helped to form a smooth EuS layer. As a final step, films were covered \textit{in situ} with $\simeq$5 nm amorphous Al$_2$O$_3$ at room temperature as a protection layer in the same deposition chamber.

In order to obtain detailed information on the crystal structure, the films were investigated by X-ray based diffraction in addition to RHEED. A well-collimated nearly background-free beam is impinged on the sample surface and the scattered X-ray intensity is collected by a two-dimensional CCD camera. The incoming beam is diffracted by a Ge (220) 4-bounce crystal monochromator to get Cu-K$_{\alpha1}$ radiation (wavelength $\lambda$=1.54056 {\AA}) over a wide range of diffraction angles. The X-ray diffraction pattern at room temperature shows two major Bragg peaks in Fig. \ref{fig:EuS_rheed_xray}(e). The more intense peak corresponds to the substrate, while the less intense one at around $\simeq$ 29.5$^\circ$ corresponds to Bragg reflection from the 5 nm EuS ([002]$_p$) layer indicating that the substrate surface is parallel to the grown layer in the STO(001)$_p$//EuS(001)$_p$ orientation. From this measurement we determine the (001)$_p$ lattice constant to be 6.06 {\AA}, which indicates a strain of $\simeq$ 2\% from the known bulk lattice constant of 5.94 {\AA}. Laue oscillations also occur near the layer's Bragg peak, which again indicates sharp surface/interface coherency. From these Laue oscillations, we can calculate the thickness ($\approx$5 nm) of the grown layer, which matches quite well to the thicknesses monitored by the quartz crystal sensor during the growth.

\begin{figure}
\centering
\includegraphics[scale=.7]{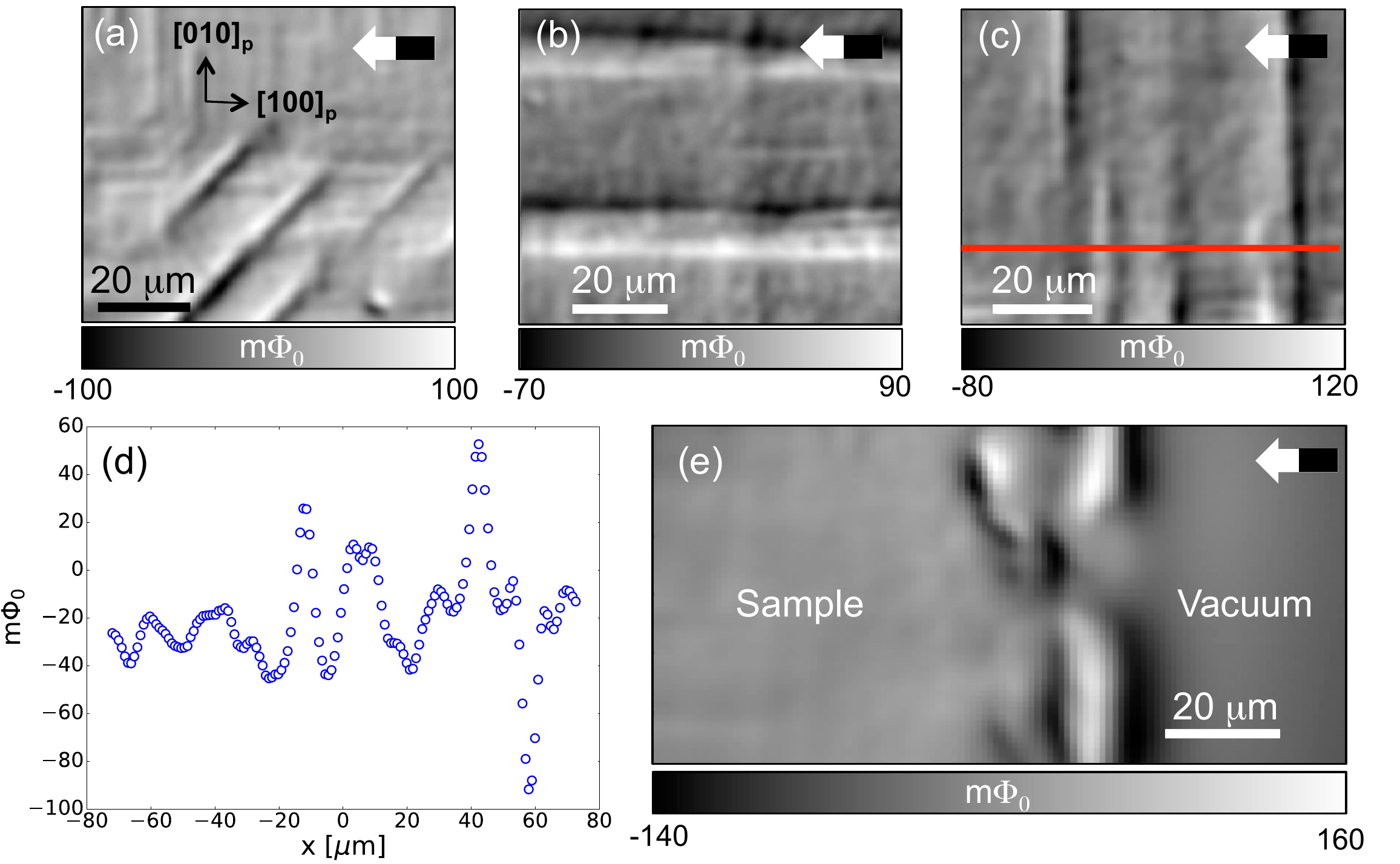}\caption{Representative magnetometry images of EuS grown on STO showing modulated magnetic features that point primarily along the [110]$_p$ (a), [010]$_p$ (b), [100]$_p$ (c) STO crystallographic axes. The black and white arrows show the direction of the training field. A line cut of image (c), shown as the red line, is plotted in (d). The edge of the sample is shown in (e), allowing us to determine the STO pseudocubic axes that are labeled in (a).}

\label{fig:EuS_mag_images}
\end{figure}

We present representative magnetometry images (Fig. \ref{fig:EuS_mag_images}(a)-(c)) of the spatial modulation of ferromagnetism (Fig. \ref{fig:EuS_mag_images}(d)) on the micron scale when the heterostructure was trained (cooled through the Curie transition) with a 13 Gauss in-plane field. Note that the amplitude of the training field does not affect the features (see appendix B). We defined the growth direction to be [001]$_p$. Unless indicated otherwise, the field training direction was pointed along the STO [$\bar{1}$00]$_p$ axis with respect to the original STO cubic crystal axes. The modulated ferromagnetic landscape appears as magnetic striations oriented along the [110]$_p$ direction (Fig. \ref{fig:EuS_mag_images}(a)), the [010]$_p$ direction (Fig. \ref{fig:EuS_mag_images}(b)), and along the [100]$_p$ direction (Fig. \ref{fig:EuS_mag_images}(c)). The crystal directions were determined by knowledge of the growth direction and calibrated by scanning the edge of the sample pointing along the [010]$_p$ direction (Fig. \ref{fig:EuS_mag_images}(e)). These magnetic striations are consistent with the expected direction of twin planes in STO because the twin planes that intersect with the surface are along the [100]$_p$, [110]$_p$, [1$\bar{1}$0]$_p$, and [010]$_p$ directions. 

We quantitatively compared the measured magnetic flux at the edge of the sample with the expected magnetic flux for a given EuS magnetization density. The mean magnetic flux magnitude of a line cut through the edge of the sample was $\approx$80 m$\Phi_0$, where $\Phi_0 = \frac{h}{2e}$ is the superconducting flux quantum. To compute the expected magnetic flux, we first compute the magnetization density. Taking the parameters of the EuS to be 28 $\mu_B$/unit cell with a lattice constant of 0.59 nm \cite{mauger1986magnetic} and a film thickness of 5 nm, we calculate an expected flux signal of approximately 200 m$\Phi_0$ (see appendix C), which is substantially larger than our measured signal at the edge (Fig. \ref{fig:EuS_supp_quant_edge}). We attribute this difference to domain structure on smaller length scales than our spatial resolution, which reduces the total signal.

To determine the relative size of the observed magnetic modulation, we compared the peak-to-peak flux signal of the features to the flux signal at the edge of the sample. Taking a line cut of Fig. \ref{fig:EuS_mag_images}(c) (shown in Fig. \ref{fig:EuS_mag_images}(d)), we measured the mean peak-to-peak magnetic flux of the modulation for 5 peaks to be $\approx$55 m$\Phi_0$ through the SQUID pickup loop. This magnetic flux is $\approx65\%$ of the flux signal at the edge, indicating that the observed spatial ferromagnetic variations are substantial. 

\begin{figure}
\centering
\includegraphics[scale=.65]{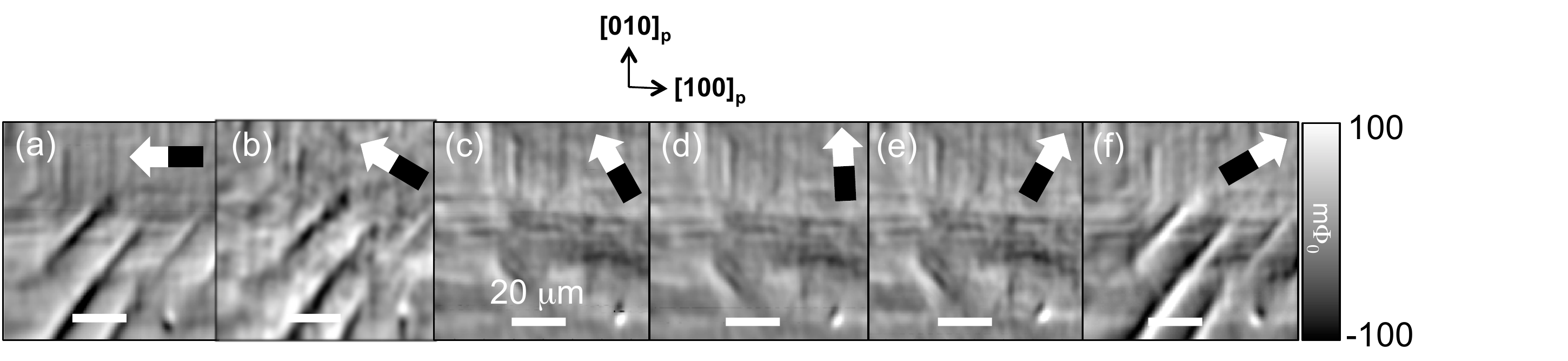}
\caption{Magnetometry dependence on in-plane field training angle. Before acquiring each image, we thermal cycled the sample to 30 K, well above the Curie temperature of EuS, and cooled with a 13 Gauss field in the direction specified by the arrow. With respect to the [100]$_p$ direction, the angle of the training field is (a) 180$^\circ$, (b) 150$^\circ$, (c) 120$^\circ$, (d) 90$^\circ$, (e) 60$^\circ$ and (f) 30$^\circ$.}
\label{fig:EuS_angle_dep}
\end{figure}

The training field polarizes some of the spins along a specific direction, and to characterize how the spin polarization direction affects the formation of these magnetic features, we examined the dependence of the magnetic configuration on the in-plane training field direction (Fig. \ref{fig:EuS_angle_dep}). Before acquiring each image, we thermal cycled the sample to 30 K, well above the expected Curie temperature of 16 K, and retrained the sample in a 13 Gauss field with different in-plane orientations. The magnetic features parallel to the [100]$_p$ and [010]$_p$ directions were independent of training-field direction, but the large magnetic features parallel to the [110]$_p$ direction disappeared for certain angles. This observation shows that the spin polarization orientation plays a role in the formation or visibility of these features.

\begin{figure}
\centering
\includegraphics[scale=.52]{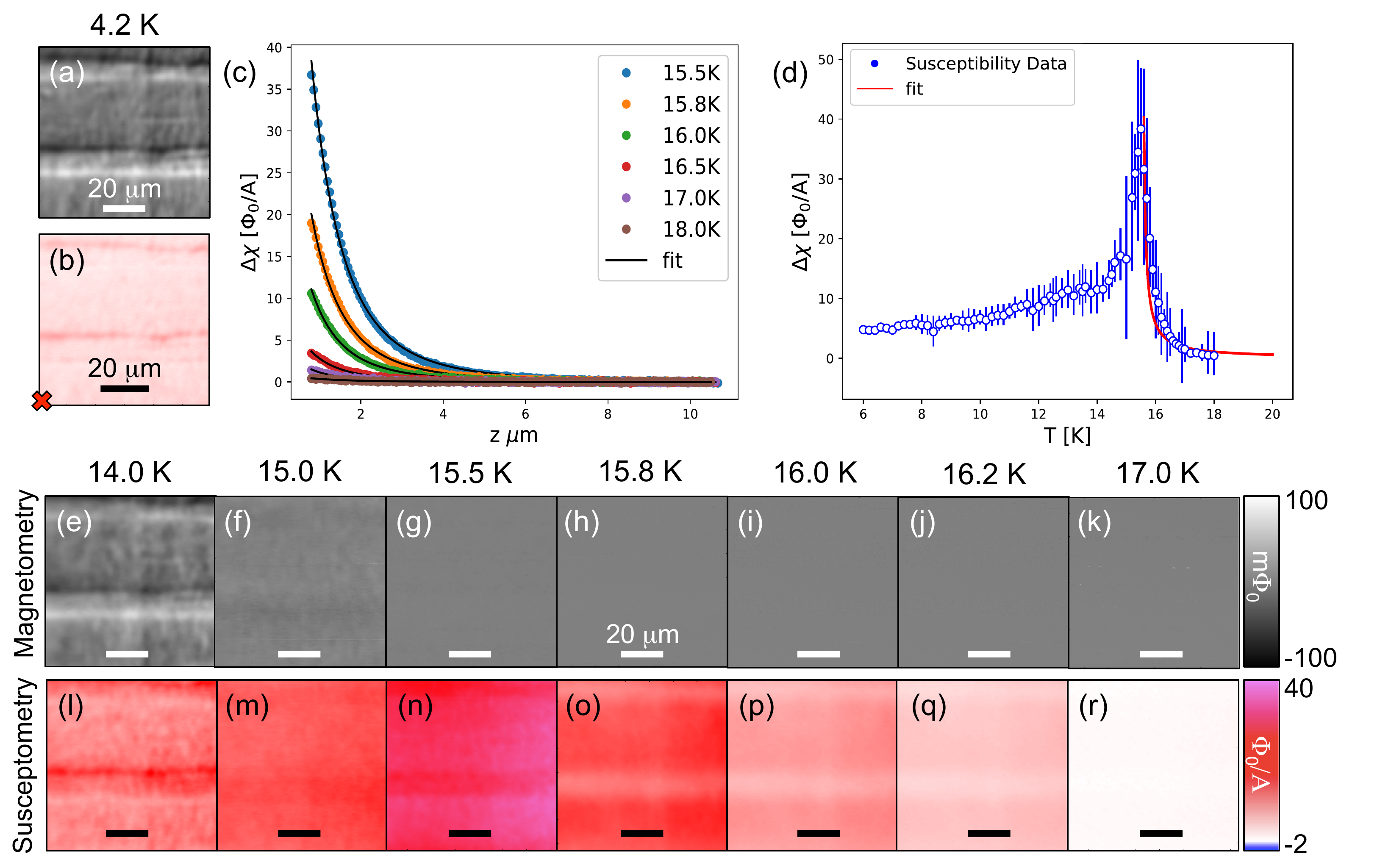}
\caption{Temperature dependence of the ferromagnetism and the susceptibility to determine the Curie temperature $T_C$. (a) Magnetometry image reproduced from Fig. \ref{fig:EuS_mag_images}(b), plotted with the corresponding (b) susceptibility image at 4.2 K. We performed susceptibility touchdowns as a function of temperature at the location in (b) marked with a red cross. (c) Representative touchdowns showing how the susceptibility changes with temperature close to and away from $T_C$. The height $z$ is defined as the separation between the SQUID substrate and the sample surface. (d) Plot of the fitted susceptibility at z = 0 with temperature, including error bars as determined by bootstrapping. The susceptibility diverges, indicative of a paramagnetic-to-ferromagnetic phase transition. The paramagnetism above the Curie temperature is fit to a Curie-Weiss law with T$_C$ = 15.4 K. (e) - (k) Magnetometry and (l) - (r) susceptibility images as a function of temperature. The paramagnetism above the fitted Curie temperature is spatially inhomogeneous, similar to the large ferromagnetic features below the Curie temperature at 4.2 K.}

\label{fig:EuS_temp_dep}
\end{figure}

To order to determine the Curie temperature $T_C$, we applied a 500 $\mu$A ac current ($f = 514$ Hz) to the field coil and performed a series of touchdowns to quantify the change in susceptibility with temperature. We performed touchdowns on the red cross (Fig. \ref{fig:EuS_temp_dep}(a) and Fig. \ref{fig:EuS_temp_dep}(b)), and we observed the susceptibility diverge as the temperature approached 15.5 K, indicating the ferromagnetic to paramagnetic transition (Fig. \ref{fig:EuS_temp_dep}(c)). When we plotted the measured susceptibility when the SQUID was closest to the sample against temperature (Fig. \ref{fig:EuS_temp_dep}(d)), we detected a divergence around 15.5 K associated with a ferromagnetic-to-paramagnetic transition. The error bars in Fig. \ref{fig:EuS_temp_dep}(d) were determined by fitting bootstrapped touchdown data shown in Fig. \ref{fig:EuS_temp_dep}(c) using the form for the susceptibility of a thin isotropic and monodomain paramagnet \cite{kirtley2012scanning} (see appendix D). The touchdown fits without bootstrapping shown in Fig. \ref{fig:EuS_temp_dep}(c) clearly deviated from the measured data, perhaps because of uncertainty in the height calibration or unaccounted piezoelectric drift. The susceptibility for T $>$ T$_C$ measures the paramagnetic response of the sample, which we fit to the Curie Weiss law $\chi(T) = \frac{C}{T - T_C}$ (Fig. \ref{fig:EuS_temp_dep}(d)), where T$_C$ is the Curie temperature and C is a proportionality constant, with T$_C$ and C as the only free parameters. The Curie-Weiss law did not perfectly capture the data (Fig. \ref{fig:EuS_temp_dep}(d)), perhaps due to height uncertainty and possible fluctuations close to T$_C$ \cite{idzuchi2014critical,als1971critical,als1976neutron}, but it does find the divergence-like peak in the paramagnetism. However, this fit yielded a fitted T$_C$ = 15.4 K, which is similar to previously reported values for EuS \cite{mauger1986magnetic}. As a check on our fitted T$_C$, we imaged the ferromagnetism with temperature (Fig. \ref{fig:EuS_temp_dep}(e)-(k)) and found that the signal became indistinguishable from noise around 15.5 K, consistent with no ferromagnetic order. Images of the paramagnetism (Fig. \ref{fig:EuS_temp_dep}(l)-(r)) above the Curie temperature showed similar spatially varying features as the ferromagnetism, suggesting that the paramagnetism is modified by a similar mechanism as the ferromagnetism.

\begin{figure}
\centering
\includegraphics[scale=.52]{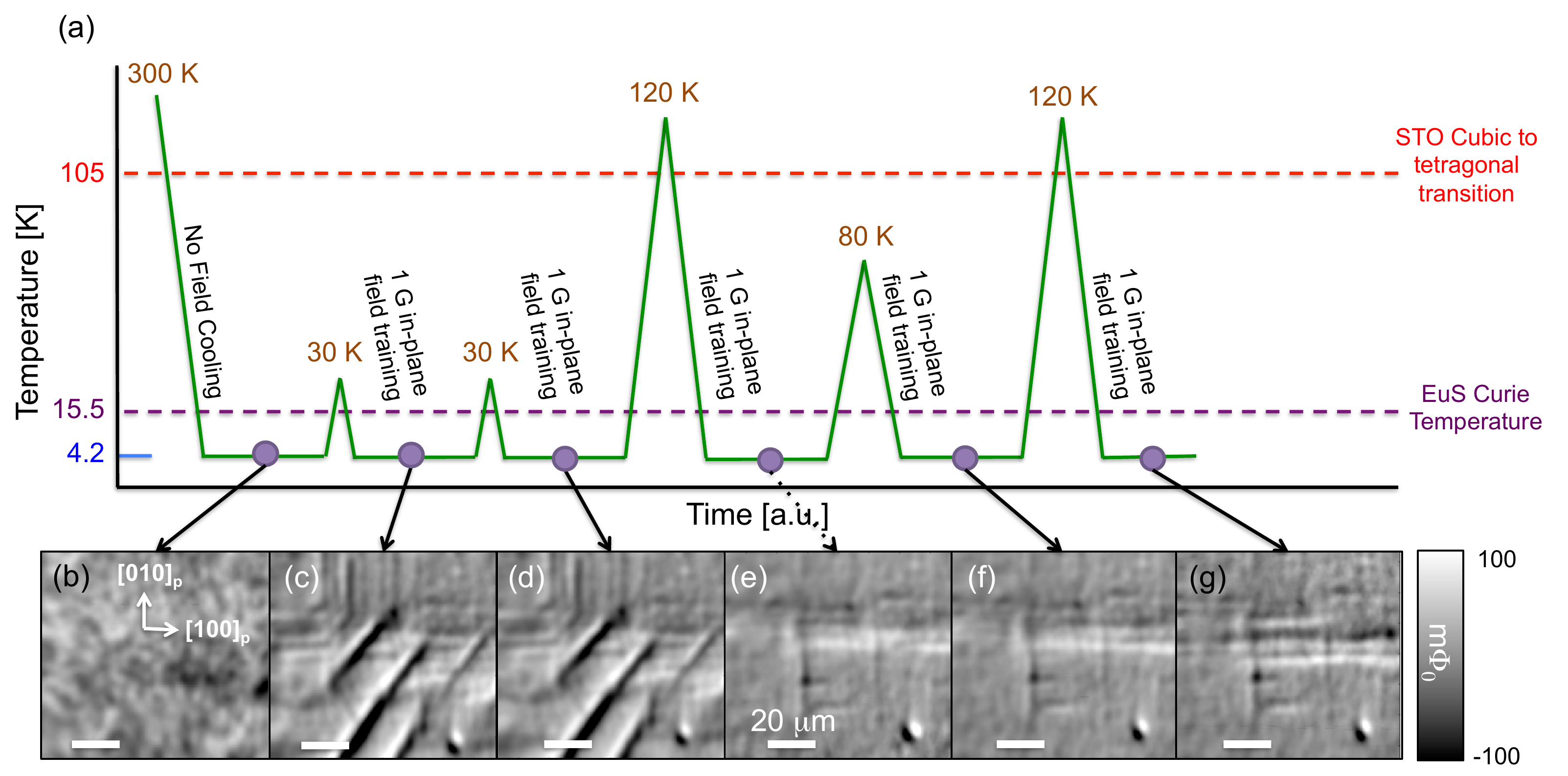}
\caption{Ferromagnetic spatial variations upon thermal cycling reveal a relationship between the magnetic structure and the STO structural phase transition. (a) Thermal history of one region; all images were taken at 4.2 K. (b) The sample was cooled from room temperature without field training, showing resolution-limited domains. (c) The sample was warmed above the Curie temperature of the EuS (30 K) and field trained with 1 Gauss, revealing a stripe-like magnetic configuration. The sample was then thermal cycled to (d) 30 K, (e) 120 K, (f) 80 K, and (g) 120 K, each cooled with a 1 Gauss field. From (d) to (e) and from (f) to (g), there are changes in the magnetic configuration corresponding to thermal cycling above the STO cubic-to-tetragonal phase transition, but the magnetic configuration does not change otherwise suggesting that the observed magnetic configuration is related to the STO structural phase transition.}
\label{fig:EuS_thermal_cycle}
\end{figure}

To demonstrate that the observed magnetic behavior is due to the STO tetragonal structure, we studied how the ferromagnetic configuration changed with thermal history (Fig. \ref{fig:EuS_thermal_cycle}(a)). Without any field training, the magnetometry image shows resolution-limited magnetic domains (Fig. \ref{fig:EuS_thermal_cycle}(b)), and we observe the magnetic striations when training in a small field (Fig. \ref{fig:EuS_thermal_cycle}(c)). When thermal cycling above the Curie temperature to 30 K (Fig. \ref{fig:EuS_thermal_cycle}(d)), the magnetic configuration does not have any distinct changes suggesting that the magnetic configuration is predetermined even before the EuS becomes ferromagnetic. However, when thermal cycling above 105 K, the magnetic configuration was substantially modified (compare Fig. \ref{fig:EuS_thermal_cycle}d to Fig. \ref{fig:EuS_thermal_cycle}(e) and Fig. \ref{fig:EuS_thermal_cycle}(f) to Fig. \ref{fig:EuS_thermal_cycle}(g)). Similar to the magnetic features in Fig. \ref{fig:EuS_mag_images}, all the features observed from thermal cycling point only along directions that intersect with twin planes, namely [100]$_p$ and [010]$_p$. Because the features only changed when thermal cycling above the STO cubic-to-tetragonal transition, and because the features only pointed along twin plane directions, we conclude that the configuration of the EuS magnetism is coupled to the configuration of the STO twin structure.

We now discuss possible origins for the observed features. One possible explanation is that the modulated magnetism is due to topography on the surface of STO from the twin planes. Scanning single electron transistor studies have shown that the STO structural transition causes the twin boundaries between tetragonal domains to have a topographical kink with a change in slope of $\tan \alpha = \frac{1}{1000}$ \cite{honig2013local}. For larger boundaries between tetragonal domains, such as 15 - 20 $\mu$m, that kink can lead to topological variations as great as 5 nm \cite{honig2013local}. Because the EuS is epitaxially grown on STO, the EuS could also experience this topography. The SQUID would measure a spatially dependent magnetic flux due to the height variation. We simulate the spatial magnetic flux from this topographical variation and find that this effect produces a spatially dependent magnetization that is two orders of magnitude smaller than the measured magnetic spatial variations (see appendix C). We conclude that surface topography does not fully explain our results.

The coupling between the EuS magnetism and the STO tetragonal structure may be magnetoelastic in origin. EuS may experience a spatially dependent strain due to STO twin formation that may alter the magnetic anisotropy along the twin boundaries giving rise to the features observed in this work. Magnetic anisotropy is determined by magnetocrystalline energy and magnetoelastic energy, and we can compare these two energy scales to determine which has a stronger influence on the anisotropy. The magnetocrystalline energy for a cubic system is expressed as $E_{mc} = K_{mc}(\alpha_1^2 \alpha_2^2 + \alpha_2^2 \alpha_3^2 + \alpha_3^2 \alpha_1^2)$ where $K_{mc} = 2.4 \times 10^4$ erg/cc and $\alpha_i$ are the directional cosines of the magnetization \cite{franzblau1967magnetocrystalline}. The magnetoelastic energy is expressed as $E_{me} = - K_{me} (x, y) \alpha_k \alpha_l$ where $K_{me} = \frac{3}{2} \lambda_{ijkl} \sigma_{ij} (x, y)$ and $\lambda_{ijkl}$ is the magnetostriction coefficients, $\sigma_{ij} (x, y)$ is the stress from the STO twins in the x - y plane. Along the [100]$_p$ direction, the highest value $K_{me} (x,y)$ is when the strain is along the [100]$_p$ direction, which is expressed as $K_{me, max} (x, y) = \frac{3}{2} \lambda_{[100]_p} \sigma_{[100]_p} (x, y) = \frac{3}{2} \lambda_{[100]_p} c_{11} \epsilon_{[100]_p} (x, y)$ where $c_{11}$ is the elastic modulus along the [100]$_p$ direction and $\epsilon (x, y)$ is the strain. Using values from the literature \cite{argyle1968magnetoelastic, liu2014elasticity}, $\lambda_{[100]_p} \simeq 10^{-5}$  and $c_{11} \simeq 120$ GPa, so $K_{me, max}(x, y) = 1.8 \times 10^7 \epsilon (x, y)$ erg/cc. Note that $\lambda_{[100]_p}$ was experimentally determined for EuO, a related Eu chalcogenide. 

To complete this calculation, we need to know $\epsilon (x, y)$. It is worth noting that X-ray diffraction measurements in Fig. \ref{fig:EuS_rheed_xray}(e) show that the out-of-plane lattice constant for 5 nm of EuS grown on STO has a strain of $\epsilon = 2 \times 10^{-2}$ at room temperature. However, this measured strain neglects how strain may vary spatially from twin formation, and calculating or experimentally determining the amount of strain for a thin film spatially is difficult. Instead, we remain agnostic as to how much strain is actually being applied, so we compare the magnetoelastic energy with the magnetocrystalline energy for a variety of strains, as shown in Table \ref{tab:strain}. For small strains, the magnetocrystalline energy dominates, so the magnetoelastic energy will have a negligible affect on the anisotropy. However, for large strains that may occur from twin boundaries, the magnetoelastic energy will strongly influence the anisotropy, which may give rise to the spatially varying magnetic structure observed in this work. 

\begin{table}
\centering
\begin{tabular}{c | c | c}
	\hline
	$\epsilon_{[100]_p}$ & $K_{me, max} (x, y)$ [10$^4$ erg/cc] & $K_{me, max}/K_{mc}$ \\ \hline
    10$^-5$ & $\simeq$ 0.018 & 0.0075 \\ \hline 
    10$^-4$ & $\simeq$ 0.18 & 0.075 \\ \hline 
    10$^-3$ & $\simeq$ 1.8 & 0.75 \\ \hline 
    10$^-2$ & $\simeq$ 18 & 7.5 \\ \hline 
	\hline
\end{tabular}
\caption{For different strains $\epsilon$ we can compare the magnetoelatic energy $K_{me, max} (x, y)$ that can vary in space with the magnetocrystalline energy $K_{mc}$.}
\label{tab:strain}
\end{table}

Strain from twin formation in STO has been observed to alter the magnetic properties of La$_{2/3}$Ca$_{1/3}$MnO$_3$ (LCMO) and La$_{0.7}$Ca$_{0.3}$MnO$_3$ (LSMO) grown on STO, perhaps also arising from a magnetoelastic origin. Bulk magnetic measurements on LCMO grown on STO \cite{ziese2008coupled} and LSMO grown on STO \cite{wahlstrom2017twinned} show changes in magnetization when cooled below the 105 K transition temperature, and micron-scale magnetic features similar to the ones reported here in Fig. \ref{fig:EuS_mag_images} were observed via magneto-optic measurements \cite{vlasko2000direct}. The conclusion of that work is that strain from the STO twin structure causes a small out-of-plane rotation of the magnetic moments, which produces the observed spatial features in magnetism. However, this explanation is not necessarily applicable to this work because the origin of the magnetism in LCMO and EuS are different (respectively, double exchange\cite{ghosh1998critical} vs. indirect exchange \cite{goncharenko1998ferromagnetic}), and we have no evidence that the EuS magnetic moments have an out-of-plane component.

Much more information is needed to confirm a magnetoelastic argument. First, one could perform X-ray diffraction measurements as a function of temperature through the cubic to tetragonal phase transition to see how the EuS lattice constant changes, although these changes will be over a large length-scale so it would still be hard to map those measurements to the micron-scale SQUID results. Second, one could apply controlled uniaxial strain to the EuS/STO heterostructure and measure how the magnetism changes with a scanning SQUID or magneto-optic technique. Finally, density functional theory calculations could shed light on how a compressed lattice constant affects magnetism in EuS.

In conclusion, here we have shown that the STO tetragonal structure modifies the magnetism of an epitaxially coupled thin film of EuS. Understanding how structural changes influence magnetism may shed light on the fundamental basis of exchange interactions and lead to the development of new and interesting systems. Thus, these changes and their impact on magnetism need to be considered when constructing devices that require homogeneous magnetic exchange.

This work was supported by FAME, one of six centers of STARnet, a Semiconductor Research Corporation program sponsored by MARCO and DARPA. The SQUID microscope and sensors used were developed with support from the NSF-sponsored Center for Probing the Nanoscale at Stanford, NSF-NSEC 0830228, and from NSF IMR-MIP 0957616. F.K. and J.S.M. acknowledge the support from NSF Grants No. DMR-1207469, ONR Grant No. N00014-13-1- 0301 and N00014-16-1-2657, and the STC Center for Integrated Quantum Materials under NSF Grant No. DMR-1231319. N.G. and F.K.  was supported by the STC Center for Integrated Quantum Materials under NSF grant DMR-1231319 (material growth). The authors also thank Adrian Swartz, Zheng Cui, Hilary Noad, and Eric Spanton for useful discussions as well as Sean Hart and Yuri Suzuki for assistance with the manuscript.

\newpage
\appendix 

\section{Measurements on other samples}

\begin{figure}
\centering
\includegraphics[scale=.6]{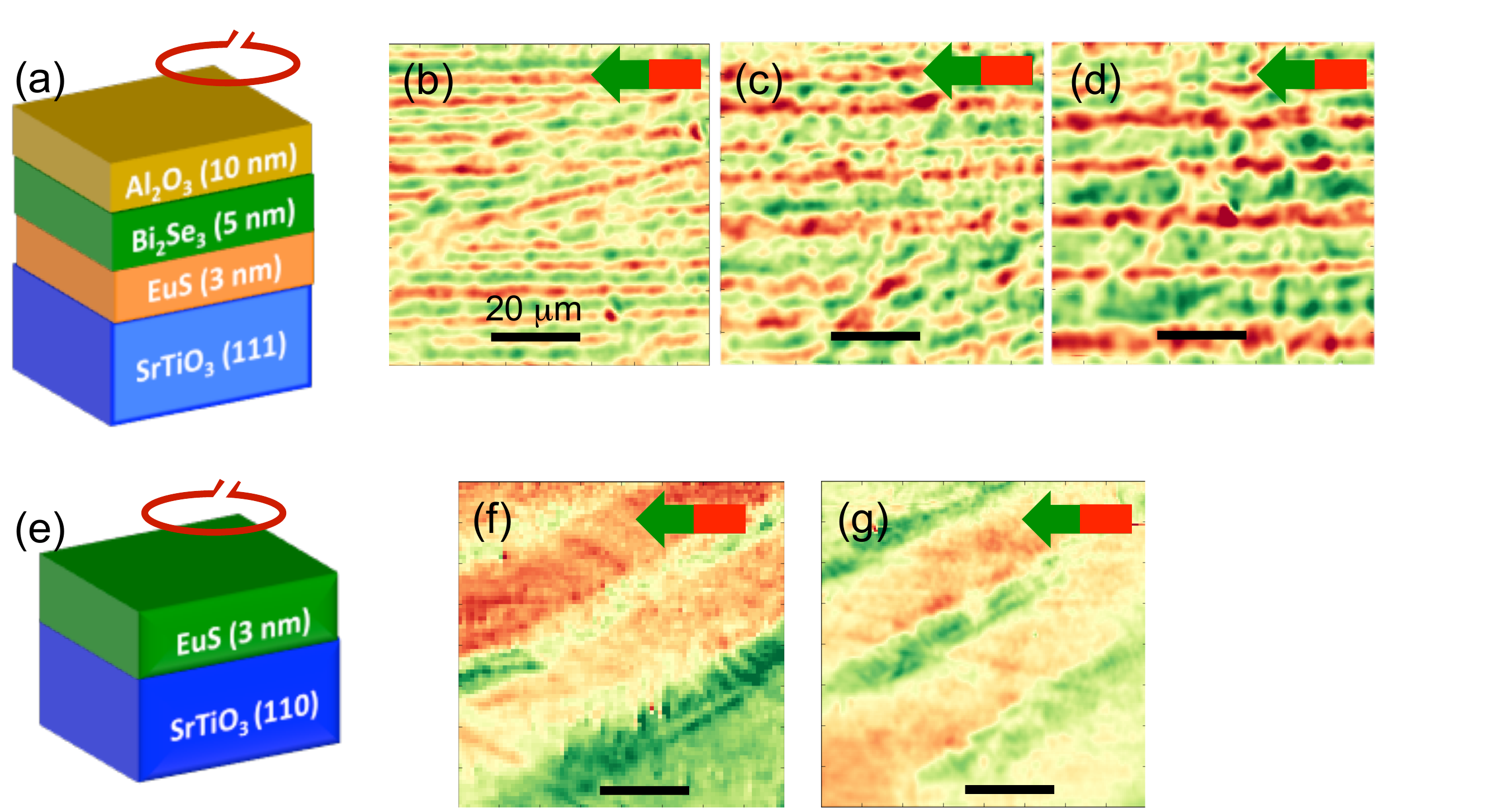}
\caption{Representative magnetic flux images for EuS grown on (111)$_p$ and (110)$_p$ oriented STO substrates showing spatially modulated magnetic flux. (a) Schematic of the EuS/STO(111)$_p$ heterostructure for magnetic flux images (b-d) that can be compared to the EuS/STO(110)$_p$ heterostructure (e) magnetic flux images (f,g). Note that unlike Fig. \ref{fig:EuS_mag_images}, the edge was not imaged so we do not know the direction of the observed features. However these features are clearly of the same nature as the features in the rest of this manuscript.}
\label{fig:EuS_supp_mag_images}
\end{figure}

The original motivation for studying this system was for studying the quantum anomalous Hall effect in Bi$_2$Se$_3$/EuS/STO films. However, we found that the EuS magnetism was spatially modulated by the STO tetragonal twin structure regardless of the STO growth plane. We show representative magnetic flux of EuS grown on (111)$_p$ oriented STO (Fig. \ref{fig:EuS_supp_mag_images}(a-d)) and (110)$_p$ oriented STO (Fig. \ref{fig:EuS_supp_mag_images}(e-g)). 

\begin{figure}
\centering
\includegraphics[scale=.55]{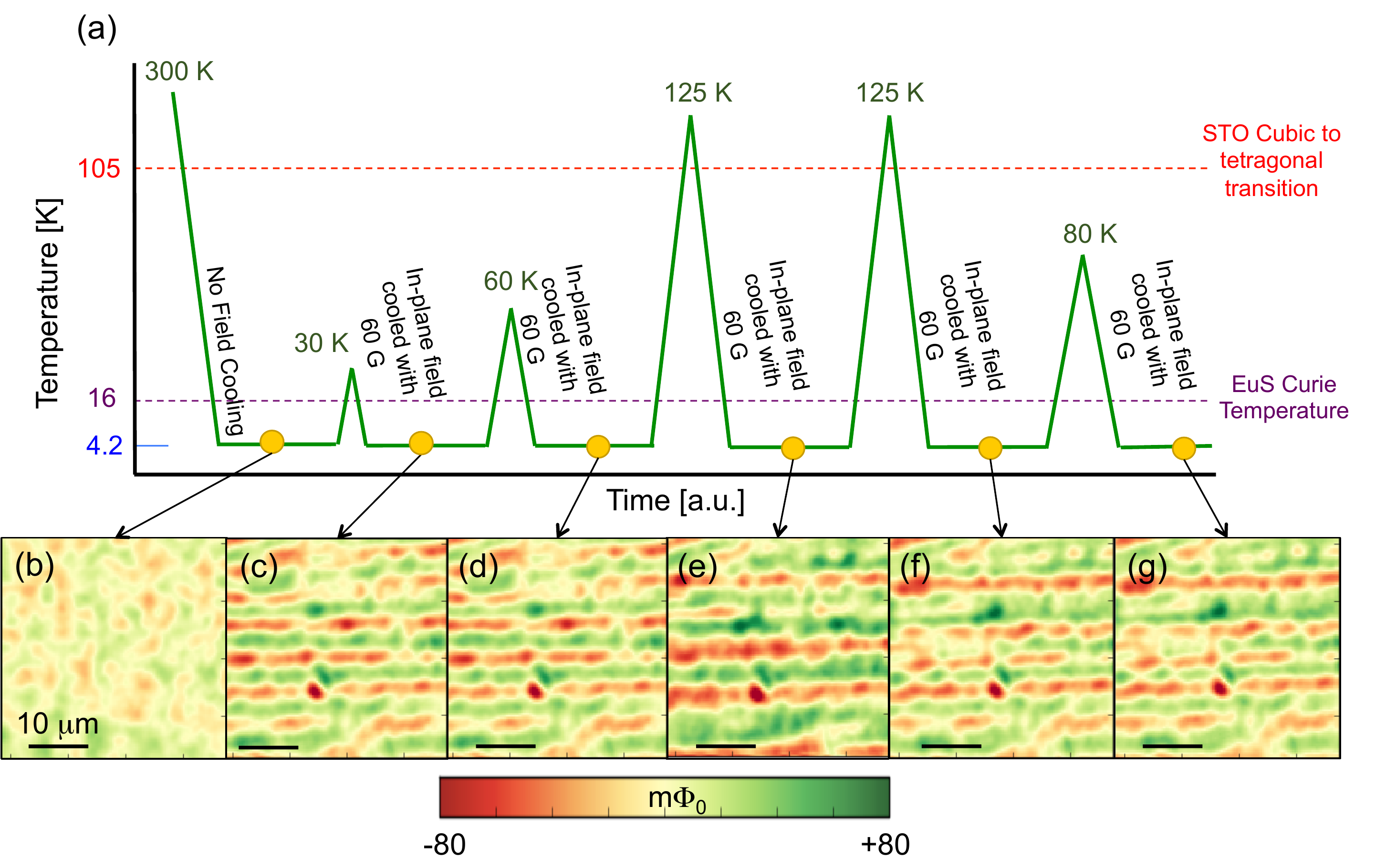}
\caption{Similar to the analysis in Fig. \ref{fig:EuS_thermal_cycle}, the ferromagnetic spatial variations are studied with thermal cycling (a) revealing a relationship between the magnetic structure and the STO structural phase transition for EuS grown on (111)$_p$ oriented STO. (b) Spatial ferromagnetism without field training shows resolution-limited domains. (c) The sample was warmed above the Curie temperature of the EuS (30 K) and field trained with 60 Gauss, revealing the spatially varying magnetic features. The sample was then thermal cycled to (d) 60 K, (e) 125 K, (f) 125 K, and (g) 80 K, each cooled with a 60 Gauss field. From (c) to (d) and (f) to (g) there are no changes in the magnetic configuration, but there are changes from (d) to (e) and (e) to (f) when thermal cycling above the STO cubic-to-tetragonal phase transition.}
\label{fig:EuS_supp_thermal_cycle}
\end{figure}

Although we do not have an image of the edge to calibrate the direction of the magnetic features, we can still confirm that they are from the STO tetragonal structure by thermal cycling (Fig. \ref{fig:EuS_supp_thermal_cycle}) for the heterostructure shown in Fig. \ref{fig:EuS_supp_mag_images}a. Thermal cycling below the STO cubic-to-tetragonal phase transition does not change the magnetic structure, but thermal cycling above does, which suggests that the EuS magnetism is coupled to the STO(111)$_p$ in this sample in the sample way that the EuS is coupled to the STO(100)$_p$ in Fig. \ref{fig:EuS_thermal_cycle}.

\section{Training field amplitude dependence of the modulated ferromagnetism}
\label{app:amp_dep}

\begin{figure}
\centering
\includegraphics[scale=.60]{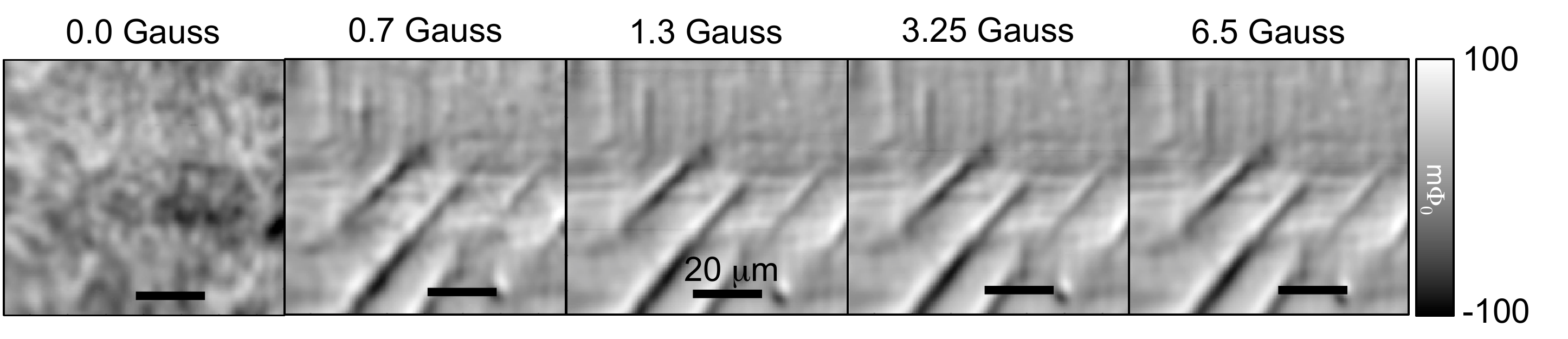}
\caption{Field training amplitude dependence on the modulated ferromagnetic features. Before each image was taken, we warmed the sample to 30 K, above the Curie temperature, and cooled it in a training field along the [$\bar{1}$00]$_p$ direction. (a) No training field is applied, showing resolution-limited magnetic domains. We observed features with (b) 0.7 G, (c) 1.3 G, (d) 3.3 G, and (e) 6.5 G. The smallest training field attempted was 0.7 G.}
\label{fig:EuS_supp_amp}
\end{figure}

We needed to apply a small training field in order to observe the modulated magnetism. Without any training field, we observed resolution-limited magnetic domains similar to SQUID measurements on conventional ferromagnets, as seen in Ref. \cite{higgs2016magnetic}. In Fig. \ref{fig:EuS_supp_amp}, we take a series of magnetic images of one 80 $\times$ 80 $\mu$m$^2$ region in which we warm the sample to 30 K, well above the Curie temperature, and cool it with a variety of training field amplitudes all pointed along the [$\bar{1}$00]$_p$ direction. The mechanism responsible for coupling the EuS magnetism to the STO tetragonal domains requires some spin polarization to be observed.

\section{Magnetic flux simulations}
\label{app:calc_flux}

To characterize the ferromagnetic features in the EuS/STO heterostructure, we simulated the magnetic flux through the SQUID pickup loop. In these simulations, each pixel in space has a fixed number of electron spins in units of the Bohr magneton $\mu_B$, and this macrospin has an in-plane angle $\phi$ and an out-of-plane angle $\theta$. The net magnetic dipole moment $\vec{m}$ in the three spatial directions is
\begin{align}
	m_x = M\cos{\phi}\sin{\theta} \\
	m_y = M\sin{\phi}\sin{\theta} \\
	m_z = M\cos{\theta},
\end{align}
where $M$ is the total number of electrons spins at each pixel. The SQUID is located at position $\vec{r}_{SQ}$, and a single dipole moment located at pixel $i, j$ is located at position $\vec{r}_{mij}$. Therefore, the distance between the SQUID and the sample is $r_{ij} = |\vec{r}_{SQ} - \vec{r}_{mij}|$ and the magnetic field from a single dipole is
\begin{equation}
\vec{B}_{dip}(\vec{r}_{ij}) = \frac{\mu_0}{4\pi}(\frac{3\vec{r}_{ij}(\vec{m}\cdot \vec{r}_{ij})}{|\vec{r}_{ij}^5|} - \frac{\vec{m}}{|\vec{r}_{ij}^3|}).
\end{equation}
Because the SQUID only measures the z-component of the magnetic field, the above equation is modified to
\begin{equation}
B_{z,dip}(\vec{r}_{ij}) = \frac{\mu_0}{4\pi}(\frac{3z(m_z \cdot \vec{r}_{ij})}{|\vec{r}_{ij}^5|} - \frac{\vec{m}}{|\vec{r}_{ij}^3|}).
\end{equation}
The measured magnetic flux for SQUID position $\vec{r}_{SQ}$ is the sum of all the B$_z$ contributions from each dipole moment:
\begin{equation}
B_{z,SQ}(\vec{r}_{ij}) = \sum_{i,j} \frac{\mu_0}{4\pi}(\frac{3z(m_z \cdot \vec{r}_{ij})}{|\vec{r}_{ij}^5|} - \frac{\vec{m}}{|\vec{r}_{ij}^3|}).
\end{equation}
We note that $\vec{r}_{SQ} = x\hat{x} + y\hat{y} + z\hat{z}$; we set $z = 1$ $\mu$m to define the height of the SQUID, and calculate B$_{Z,SQ}(x,y)$ representing a 2D magnetization density image. The final magnetic flux image is approximated by convolving B$_{Z,SQ}$ with a circular disk of radius 1.27 $\mu$m, representing the SQUID point-spread function. Mathematically this operation is expressed as $\Phi_{SQ} =\int g_{disk}(x,y)B_{Z,SQ} da$, where $g(x,y)$ represents the point-spread function, the integral is taken over the SQUID plane, and $da$ is the area vector pointing normal to the SQUID plane, which is parallel to B$_z$.

Eu has 7 $\mu_B$/atom, and in an NaCl structure, there are 28 $\mu_B$/unit cell. The unit cell has a lattice constant of 0.59 nm. For a 5 nm film in a 1 $\mu$m$^2$ pixel, there are 28 $\mu_B$*(5 nm/0.59 nm)*(1000 nm/0.59 nm)*(1000 nm/0.59 nm) = 6.81 $\times$ 10$^8$ $\mu_B$. We use this value as M for each pixel in space and assume that the orientations of all these moments are the same. From the observed magnetic structure (Fig. \ref{fig:EuS_mag_images}), clearly this assumption is not valid due to domains that probably occur on length scales much smaller than one pixel, so we expect that calculations using this method yield an upper bound of the measured magnetic flux.

\begin{figure}
\centering
\includegraphics[scale=.7]{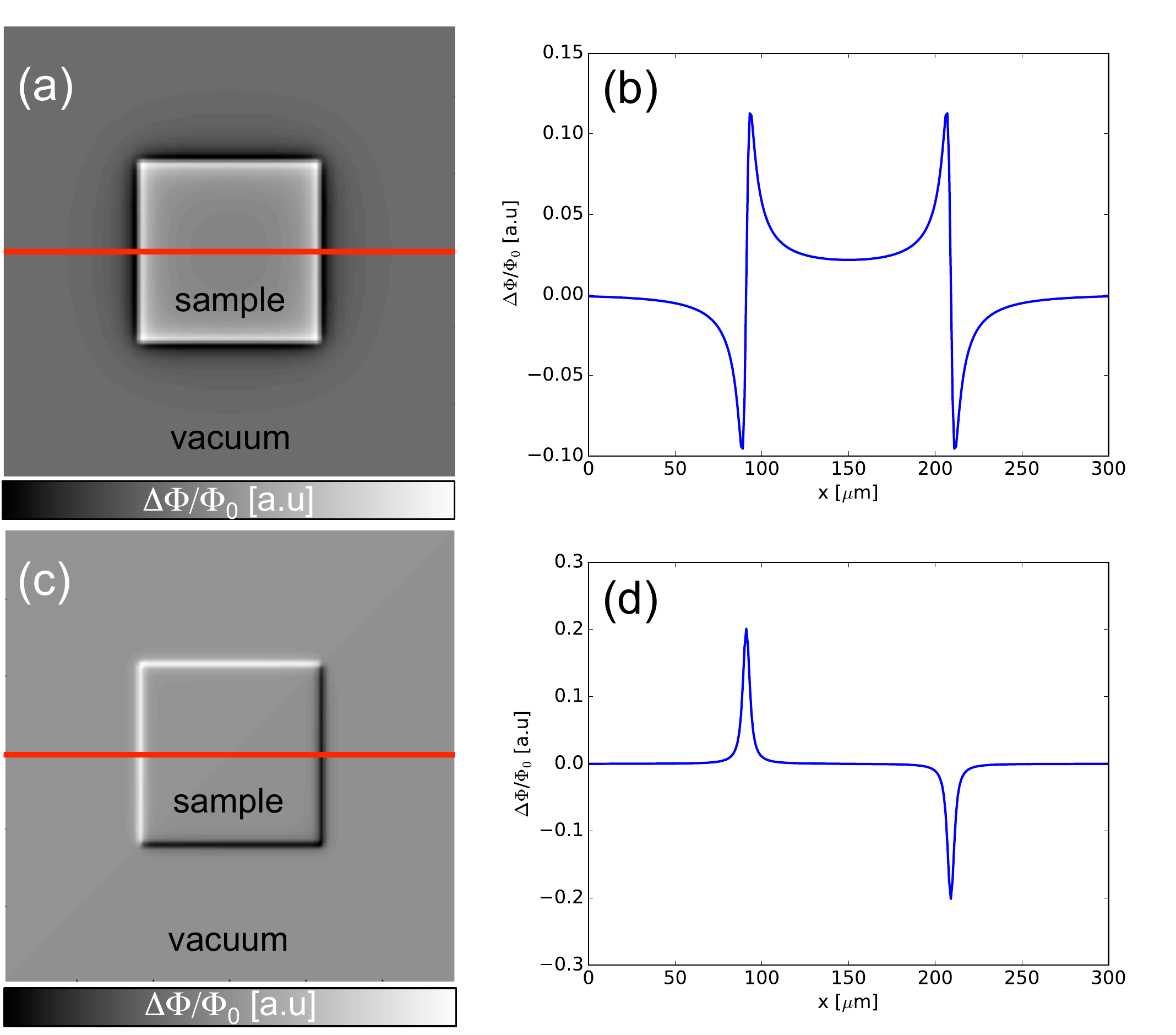}
\caption{A simulation of out-of plane magnetization (a, b) is compared to a simulation of in-plane magnetization (c, d) as seen by the SQUID. Qualitative comparisons are shown as 2D images (a, c), and as linecuts through the images (b, d). The images are 300x300 $\mu$m$^2$.}
\label{fig:EuS_supp_ip_opp}
\end{figure}

Because the SQUID only measures B$_z$, in-plane and out-of plane magnetizations show similar structure in the measured magnetizations because both have B$_z$ components. However, the difference between the magnetizations can be determined by measuring the edge of the sample. We simulated the qualitative differences between in-plane and out-of plane magnetization (Fig. \ref{fig:EuS_supp_ip_opp}). For this simulation, we assumed no domain structure to complicate the image and interpretation, and we made the sample size $100 \times 100$ $\mu$m$^2$, but in principle the same images would be produced if the simulated sample were the size of the actual sample. The difference between in-plane and out-of plane magnetizations qualitatively is that the interior of a sample with in-plane magnetization produces zero magnetic flux, while the out-of plane magnetization interior has a constant flux offset relative to the magnetization of the sample. The easiest way to distinguish the two magnetizations is by comparing the line cuts for out-of-plane (Fig. \ref{fig:EuS_supp_ip_opp}) and in-plane (Fig. \ref{fig:EuS_supp_ip_opp}d) signals, which display quite different edge behavior. 

\begin{figure}
\centering
\includegraphics[scale=.5]{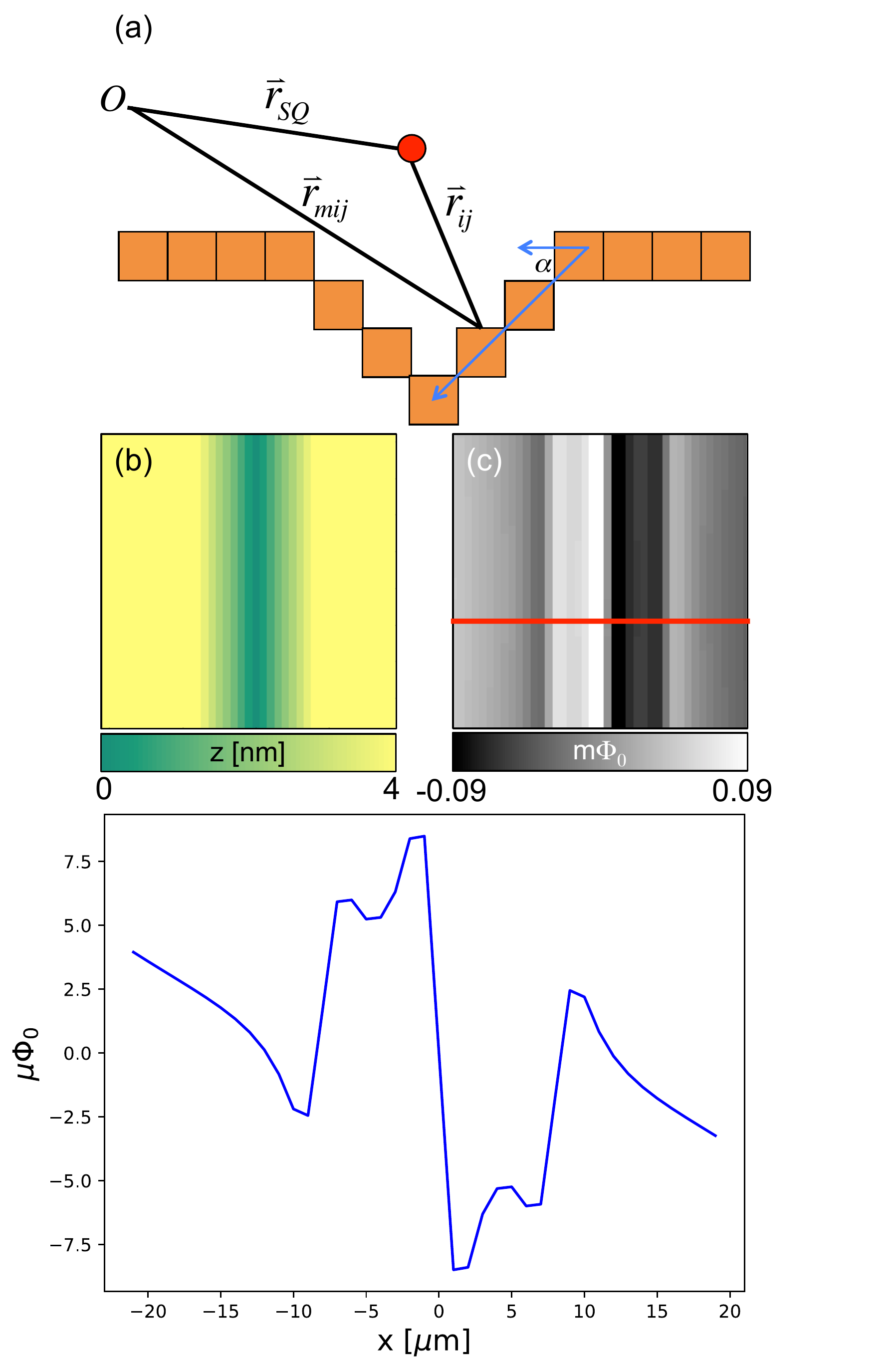}
\caption{Simulations of topographical features in magnetization produce a small spatially varying flux that cannot fully account for the observations in Fig. \ref{fig:EuS_mag_images}. (a) Cartoon of how STO topography affects the measured magnetic flux. The red dot represents the SQUID, each orange box represents a pixel of EuS magnetism, which would theoretically have $6.81 \times 10^8 \mu_B$. The angle $\tan{\alpha} =\frac{1}{1000}$ of the topography for STO \cite{honig2013local}, and is not drawn to scale. Although all spins contribute to the final measured flux, the distance $\vec{r}_{ij}$ varies spatially with respect to the SQUID distance to the origin $\vec{r}_{SQ}$ and the distance of the $i, j^{th}$ moment to the origin $\vec{r}_{mij}$. (b) Image of topography for a 18 $\mu$m twin boundary for the measured $\alpha$ showing the underlying topographical structure of the magnetization for the simulation. (c) The flux through the SQUID pickup loop given the topography in (b) for a fixed number of spins at each pixel all in the same orientation shows that the topography causes the magnetic flux to vary spatially. (d) A line cut of the simulation showing that the simulated peak-to-peak signal is ~0.2 m$\Phi_0$, which is approximately 2 orders of magnitude smaller than the measured peak-to-peak signal of 55 m$\Phi_0$ in Fig. \ref{fig:EuS_mag_images}, suggesting that topography alone cannot  explain the observed signal. All images are $80 \times 80$  $\mu$m$^2$.}

\label{fig:EuS_sim_top}
\end{figure}

As discussed in the main text, these simulations ruled out one trivial explanation for our observations: that the modulated magnetism is solely due to topography on the surface of STO due to the twin planes. We simulated such a scenario by setting $M = 6.82 \times 10^8 \mu_B$ with a fixed orientation at each point in space and defining a variable distance between the SQUID plane and the distance to the magnetic layer (see Fig. \ref{fig:EuS_sim_top}a for a cartoon of this simulation). The topographical stripe occurs over 18 $\mu$m, resulting in a $\approx$4.5 nm topography (Fig. \ref{fig:EuS_sim_top}b). The final simulation of the magnetic flux from a magnetic film due to a shallow topography showed that the calculated magnetic flux does vary spatially (Fig. \ref{fig:EuS_sim_top}c, line cut in Fig. \ref{fig:EuS_sim_top}d), but the signal is two orders of magnitude lower than the measured magnetic signal (Fig. \ref{fig:EuS_mag_images}). Based on these simulations, we conclude that topography on the surface of STO cannot fully explain our observations. 

\begin{figure}
\centering
\includegraphics[scale=.75]{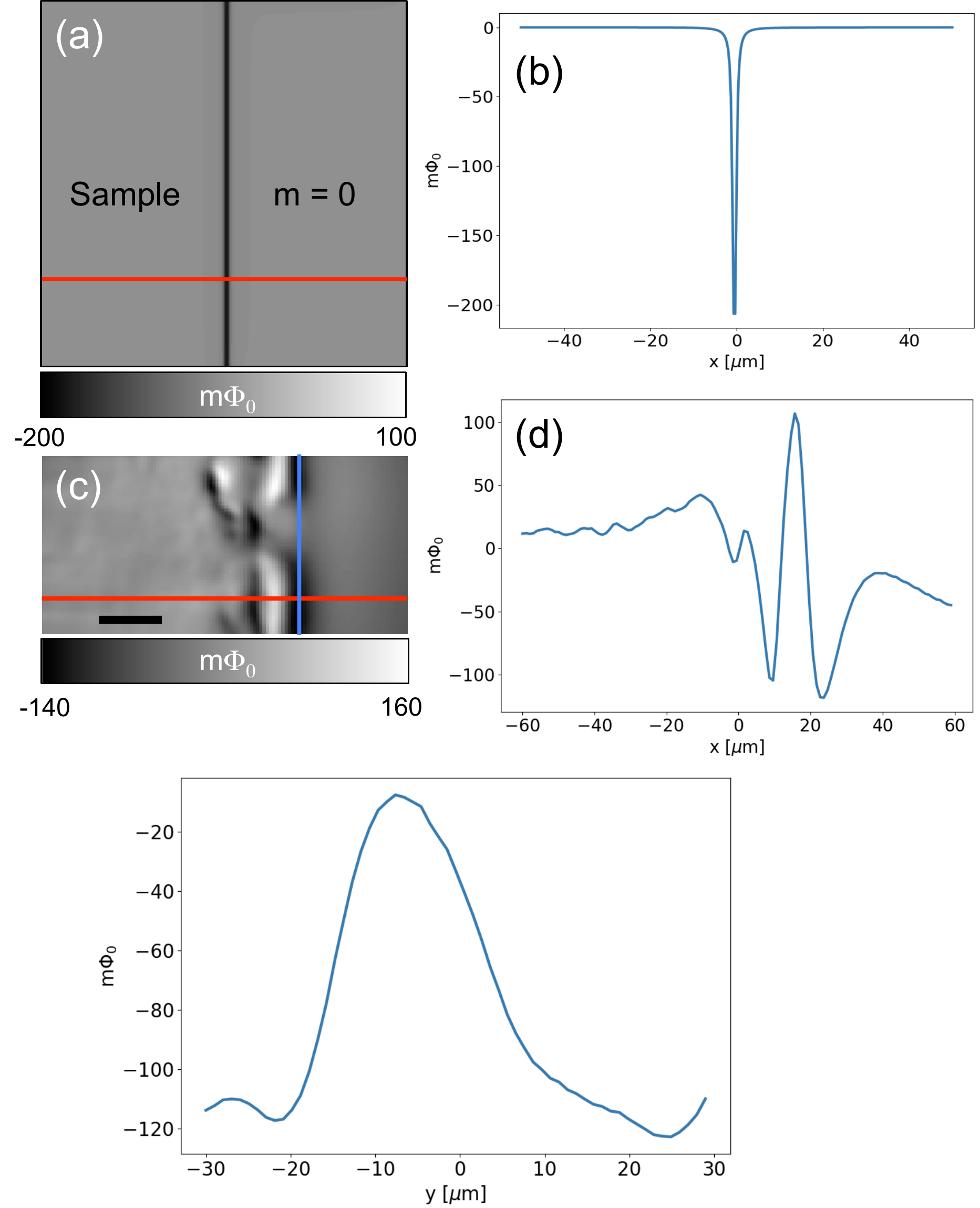}
\caption{(a) Magnetic flux simulation of an edge with the magnetic moment at each pixel equal to that of EuS. Values of the red line are shown in (b). Reproduced image of the edge in Fig. \ref{fig:EuS_mag_images}e. The red line is a horizontal line cut (d), and the blue line is a vertical line cut (e).}
\label{fig:EuS_supp_quant_edge}
\end{figure}

We compared the measured magnetic flux to the expected magnetic flux for calculated magnetization density for EuS (Fig. \ref{fig:EuS_supp_quant_edge}). The computed flux signal at the edge is $\simeq$ 200 m$\Phi_0$. To keep the sign of the magnetic flux at the sample edge consistent between the calculation and the measurement, we set M to be negative, which is why Fig. \ref{fig:EuS_supp_quant_edge}b has a downward peak. The vertical line cut of the magnetic flux at the edge reveals a mean value of 82 m$\Phi_0$ (Fig. \ref{fig:EuS_supp_quant_edge}e). The 'missing' flux at the edge of the sample may be due to defects in the sample, so the comparison average value of the measured magnetic flux is approximately 120 m$\Phi_0$, which is still less than the computed 200 m$\Phi_0$ at the edge.

\begin{figure}
\centering
\includegraphics[scale=.65]{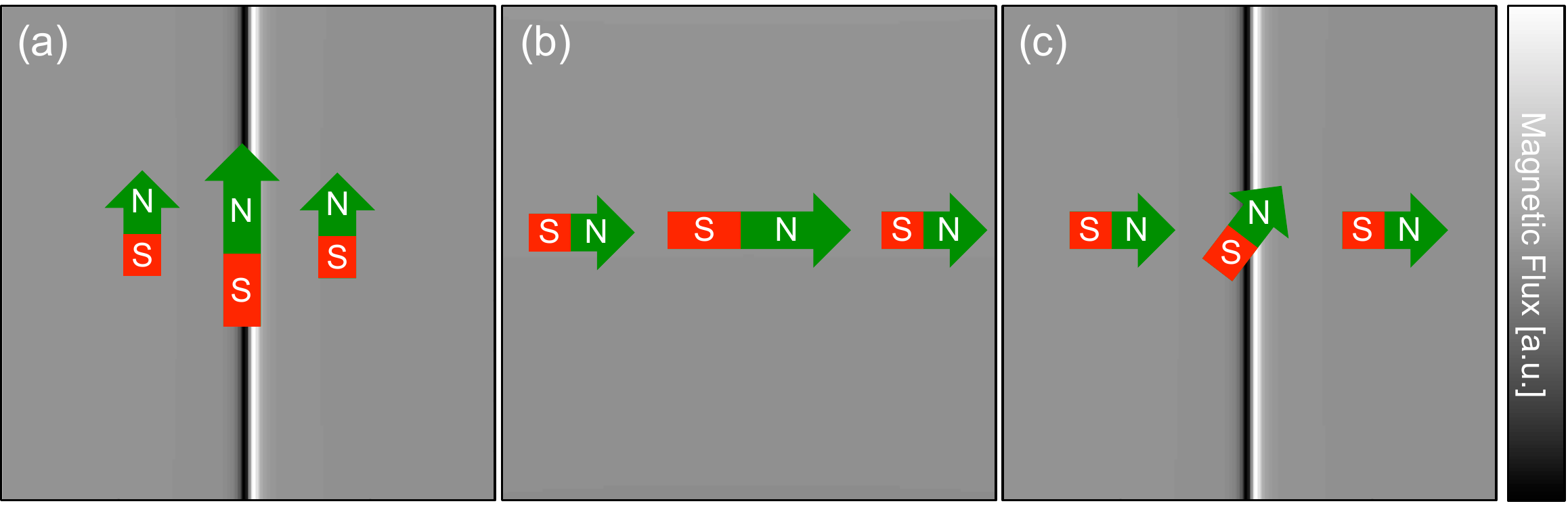}
\caption{Simulations showing possible magnetic configurations for spins on and off of twin boundaries that produce similar features found in Fig. \ref{fig:EuS_mag_images}). (a) Simulations showing one possible macrospin configuration in which there are more spins on than off a vertical stripe representing a twin boundary. This simulation shows magnetic flux features similar to the data in Fig. \ref{fig:EuS_mag_images}. However, when the spins point perpendicular to the vertical stripe (b), the signal disappears, which is in conflict with the observation in Fig. \ref{fig:EuS_angle_dep}. (c) Another possible macrospin configuration in which the spins on a vertical stripe are canted with respect to the spins away from a stripe. The resulting magnetic flux simulation displays features that are qualitatively similar to the features in Fig. \ref{fig:EuS_mag_images}. All images are $80 \times 80$ $\mu$m$^2$.}

\label{fig:EuS_sim_mag}
\end{figure}

Similar simulations can also shed light on the underlying magnetic structure that results in our observations. Because the SQUID measures a scalar flux, we cannot a priori reconstruct the components of the 3D magnetization of the sample. However, we simulated distinct possibilities and compare the results qualitatively against the data. Although here we do not propose a physically motivated mechanism, the simulations that most resemble the data involve no out-of-plane moment (Fig. \ref{fig:EuS_sim_mag}), consistent with measurements of thin films of EuS \cite{wei2013exchange,katmis2016high} and consistent with the in-plane magnetization at the edge of the sample (\ref{fig:EuS_mag_images}e). The first configuration that qualitatively matches the data is one in which there are more spins, and thus a higher magnetization, on the twin boundaries, as represented by a stripe in Fig. \ref{fig:EuS_sim_mag}a. However, we ruled out this scenario because when the spins are oriented parallel to the stripe, the simulation shows that there is no measured magnetization (Fig. \ref{fig:EuS_sim_mag}b), which is inconsistent with the results in Fig. \ref{fig:EuS_angle_dep}. The intuition behind this result is that the moments of the same size pointed in the same direction cancel out each other out. Alternatively, perhaps the spins on the stripe representing a twin boundary are slightly rotated with respect to the spins on a tetragonal domain (Fig. \ref{fig:EuS_sim_mag}c). Although this scenario qualitatively agrees with the experimental features, we do not propose a microscopic reason why EuS spins would be canted at an STO twin boundary.

\section{Quantifying paramagnetism}
\label{app:param_calc}

Applying an ac current to the field coil on the SQUID sensor allows us to locally modulate the sample magnetism. We measured the sample response through the pickup loop using standard phase sensitive detection. The susceptibility signal is interpreted as a change in mutual inductance between the pickup loop and the field coil. The lock-in voltage is converted to this mutual inductance change by
\begin{equation}
\Delta M = \frac{\Phi_0}{V_{flux}}\frac{V_{meas}}{GI_{FC}},
\end{equation}
where $M$ is the mutual inductance, V$_{meas}$ is the measured voltage, $G$ is the lock-in amplifier gain, I$_{FC}$ is the field coil current, and $\Phi_0$/V$_{flux}$ is the flux-to-voltage conversion determined by the SQUID modulation. The offset is determined by measuring the change in mutual inductance as a function of height (referred to as a touchdown) and subtracting the point of the signal at furthest spacing (approximately 20 $\mu$m), indicating no change in mutual inductance by the sample and thus zero susceptibility. Knowing that the sample is a thin film paramagnet, the functional form of the mutual inductance of the touchdown is
\begin{equation}
\Delta M(z) = \frac{Ctz}{(1+4z^2)^{5/2}},
\end{equation}
where $t$ is the thickness of the EuS, $C$ is a constant that depends on the permeability, and $z$ is the SQUID height \cite{kirtley2012scanning}. We fit the data based on this form to extract the susceptibility, and error bars were determined by bootstrapping. 

\label{appendix:param_calc}

\section{X-ray diffraction and RHEED}

Fig. \ref{fig:EuS_rheed_xray} shows the RHEED pattern of a 5 nm EuS layer immediately after growth, with the electron beam along the [110]$_p$ and [001]$_p$-azimuths (Fig. \ref{fig:EuS_rheed_xray}a, b) and after annealing (Fig. \ref{fig:EuS_rheed_xray}c, d). See main text for more details.

The crystalline quality of the thin film is studied by 1.54 {\AA} Cu - K$_{a1}$ x-ray diffraction showing two major Bragg peaks in Fig. \ref{fig:EuS_rheed_xray}e. The more intense peak corresponds to the substrate, while the less intense one at around 30$^\circ$ corresponds to Bragg reflection from the 5 nm EuS([200]) layer indicating that the substrate surface is parallel to the grown layer in the STO(100)$_p$//EuS(100)$_p$ orientation. Laue oscillations also occur near the layer's Bragg peak, which again indicates sharp surface/interface coherency. From these Laue oscillations, we can calculate the thickness ($\approx$5 nm) of the grown layer, which matches quite well to the thicknesses monitored by the quartz crystal sensor during the growth.

\bibliography{EuS_STO_bib}

\end{document}